# Spectral Bounds on Entropy and Ergotropy via Statistical Effective Temperature in Classical Polarization and Quantum Thermal States

Tariq Aziz[1], Meng-Long Song[1], Liu Ye[1], Dong Wang[1,*], José J. Gil[2,*], and Sabre Kais[3]

[1] *School of Physics and Optoelectronic Engineering, Anhui University, Hefei 230601, China*
[2] *Photonic Technologies Group, University of Zaragoza, Pedro Cerbuna 12, 50009 Zaragoza, Spain*
[3] *Department of Electrical and Computer Engineering, North Carolina State University, Raleigh, NC 27606, USA*

*Corresponding author emails:
dwang@ahu.edu.cn
ppgil@unizar.es

**Abstract**

We formulate a unified definition of the statistical effective temperature (SET), $\tau_d$, for finite-dimensional classical and quantum systems, using dimension-dependent indices of purity derived from the eigenvalue spectrum. This spectral approach bypasses the need for Hamiltonians or energy gaps and remains applicable to both quantum density matrices and classical polarization coherency matrices. The SET framework naturally describes the divergence of inverse temperature near pure, non-degenerate states, consistent with the third law. Using entropy–SET ($S_d - \tau_d$) diagrams, we explore spectral bounds in two-, three-, and four-level systems, which reveal physically realizable entropy regions, rank-dependent constraints, and cusp-like features. A Hamiltonian-free parametric spectrum ansatz provides a universal reference curve within these bounds. Furthermore, we derive spectral bounds on ergotropy as a function of entropy and SET, which quantify the maximum extractable work under passive constraints and introduce the notion of structure-states; engineered spectral configurations that saturate these bounds. Our analysis shows that SET serves as a thermodynamically meaningful and operationally relevant quantity for bounding entropy and ergotropy in both classical polarization systems and quantum thermal states.

## 1. Introduction

Temperature is one of the most fundamental concepts in thermodynamics and statistical mechanics, which is typically defined through the canonical (Gibbs) distribution in equilibrium systems, where the relative probabilities of the microstates of the system follow the Boltzmann factor. However, this definition becomes less straightforward when one moves beyond strict equilibrium conditions, as seen in finite-size quantum systems, partially coherent classical light fields, and non-thermal distributions [1-5]. In these regimes, the usual notion of temperature becomes more complex.

Several researchers have turned their attention to alternative parameters, such as an effective or virtual temperature to characterize thermodynamic-like behavior in a classical or quantum system [4-13].

The concept of effective or virtual temperatures plays an important role in the interpretation of systems that exhibit quantum thermodynamic behavior, where small quantum systems can show extreme behaviors such as negative or very high virtual temperatures that arise from population inversions and other non-equilibrium effects [7-14]. These phenomena are essential to understand work extraction, passivity, and fundamental limits in thermal machines. Specifically, when equilibrium thermodynamic frameworks are no longer sufficient to describe these behaviors [8-9]. Similarly, in classical polarization optics, the work of Brosseau and Bicout [5] introduced an effective polarization temperature to describe the loss of polarization in optical fields. They compare the two-level Ising spin system with the classical degree of polarization of a planar (2D) electromagnetic field when it is scattered through random media [5, 15]. This analogy to the Ising spin system enables them to view entropy generation during multiple scattering as a thermal-like process.

Despite these advancements, many existing models of effective or virtual temperature rely on simplified assumptions, such as pairwise transitions or two-dimensional systems, and often presuppose uniform energy gaps or specific Hamiltonian structures [5,12]. These assumptions can be limiting, especially when we deal with systems with larger Hilbert spaces, such as multi-level quantum states or classical light fields with multiple polarization components. As a result, a more general approach is needed to extend the concept of effective temperature to higher-dimensional systems.

In this work, we propose a unified, finite-dimensional approach to defining the statistical effective temperature (SET) denoted as $\tau_d$, a concept applicable to both classical polarization states and quantum thermal states. Our approach is based on two observations: first, both quantum density matrices and classical normalized polarization coherency matrices share the same mathematical properties. They are positive-semidefinite and Hermitian operators with unit trace. Second, the von Neumann entropy naturally measures the distribution of eigenvalues of such operators, and a set of indices of purity (IPs) succinctly provides the structure of these eigenvalues, which translate directly into a dimension-dependent temperature parameter. Importantly, the inverse SET denoted as $\beta_d$ we defined diverges as the system approaches a pure non-degenerate state, which aligns with the third law of thermodynamics. Our definition naturally reduces to well-established two-level (or two-dimensional) effective temperature formulations for simpler systems [5].

This approach avoids assumptions about energy gaps or pairwise transitions [12], which makes it applicable to a wide range of finite-dimensional quantum systems, classical polarization states, and multi-level systems. Moreover, we compare SET to the canonical temperature $T$ in finite isotropic Heisenberg chains (up to $L = 9$) and observe low-temperature plateaux in odd-length chains due to ground-state degeneracy. While $T \to 0$, in these cases, the corresponding $\tau_d$ remains finite, which show how SET depicts spectral disorder independently of energy-level scaling and describes its distinct sensitivity to degeneracy structure.

Furthermore, the entropy–SET ($S_d - \tau_d$) diagrams generated by this approach reveal geometric features, such as cusp points and boundary curves, which resemble phase-transition-like effects or signatures of dimensionality. These diagrams provide a robust tool to understand the thermodynamic behavior of both classical and quantum systems. To construct a reference path through the space of physically realizable states, we introduce a Parametric Spectrum Ansatz (PSA), which generates a family of spectra using a thermal-like distribution over a fixed set of dimensionless energy levels [16-19]. The PSA provides a deterministic, Hamiltonian-agnostic parameterization of spectral evolution constrained only by purity and entropy, which allows a smooth interpolation between maximally mixed and pure states, which lies inside the $S_d - \tau_d$ diagram. Additionally, we provide Hamiltonian-specific analyses for qubit (2D polarization states), qutrit (3D polarization states), and quartit (4D states represented by Mueller matrices) thermal states. These analyses demonstrate bounded entropy regions, rank-dependent constraints, and cusp points, which show characteristic features in classical polarization states and facilitate a comparison with quantum thermal states. It is observed that the thermal entropy curves with varying standard temperatures exhibit similar behavior to the $S_d - \tau_d$ diagram. For specific values of $\omega$ ($\omega_1$), the entropy curves approximately coincide with the upper boundary of the $S_d - \tau_d$ diagram, which strengthens the universality of the approach.

In addition to entropy, we further examine how SET governs energetic performance by analyzing its relation to ergotropy, the maximum extractable work under unitary operations. We assumed structured spectra, which gives operational bounds on work extraction for arbitrary quantum states and a randomly generated diagonal Hamiltonian and introduced the notion of structure-state. This relates ergotropy to both von Neumann entropy and SET. These ergotropy–entropy and ergotropy–SET diagrams, supported by extensive numerical simulations, reveal that SET provides a consistent framework for comparing energy-extraction capabilities across systems of different dimensions, even in the absence of thermal equilibrium.

The following sections of this paper formally define the SET for finite-dimensional density matrices and associated purity measures. We explore the thermodynamic implications of this work with a focus on its basis independence and validity with the third law, and we demonstrate how it recovers known results in classical polarization theory.

## 2. Statistical Effective Temperature (SET)

In this section, we define a finite-dimensional interpretation of the SET that demonstrates thermodynamic-like properties. The approach begins with a $d$-dimensional density matrix $\rho$, which is positive semidefinite, Hermitian, and normalized to unity. Let $= \{\lambda_1, \lambda_2, \ldots, \lambda_d\}$ be its descending-ordered eigenvalues, which satisfy normalization, $\sum_{i=1}^{d} \lambda_i = 1$. The von Neumann entropy of $\rho$ is given by [20]

$$S(\rho) = -\mathrm{Tr}(\rho \ln \rho), \tag{1}$$

where Tr stands for trace. Eq. (1) takes its minimum value of 0 for a pure state (rank 1) and a maximum of $\ln d$ when $\rho$ is the maximally mixed state $I_d/d$ with $I_d$ being the $d \times d$ identity matrix.

### 2.1 A single qubit in a transverse magnetic field

We begin with a quantum Ising model for a qubit state. The quantum Ising model serves as a fundamental framework in statistical mechanics and quantum thermodynamics, which provides insight into thermal and quantum fluctuations, phase transitions, and non-equilibrium dynamics [21]. We consider a two-dimensional density matrix and its eigenvalue distribution to analyze a single qubit in the presence of a transverse magnetic field. The Hamiltonian is given by

$$H = -J\sigma_z - h\sigma_x, \tag{2}$$

where $J$ represents the interaction strength, we set $J = 1$ for simplicity, $\sigma_z$ is the Pauli $z$-matrix with eigenvalues $+1$ and $-1$, and $\sigma_x$ is the Pauli $x$-matrix responsible for flipping the qubit between states $|0\rangle$ and $|1\rangle$. The parameter $h$ denotes the strength of the transverse magnetic field, which induces quantum fluctuations in the system. In thermal equilibrium, the state of the system is described by the density matrix $\rho$, which follows the Boltzmann distribution,

$$\rho = \frac{e^{-\beta H}}{Z}, \tag{3}$$

where $\beta = \frac{1}{T}$ is the inverse temperature, and $Z$ is the partition function that ensures the normalization of the density matrix and is given as

$$Z = \mathrm{Tr}(e^{-\beta H}). \tag{4}$$

The eigenvalues of the Hamiltonian $H$ can be explicitly computed. The matrix representation of the $H$ is,

$$H = -\begin{bmatrix} 1 & h \\ h & -1 \end{bmatrix}. \tag{5}$$

The eigenvalues $\alpha_\pm = \pm\sqrt{1+h^2}$ are obtained by solving the characteristic equation, $\det(H - \alpha I) = 0$, where $I$ is the identity matrix. The partition function $Z$ is the sum of the exponentiated eigenvalues of the Hamiltonian,

$$Z = e^{-\beta\sqrt{1+h^2}} + e^{\beta\sqrt{1+h^2}} = 2\cosh\left(\beta\sqrt{1+h^2}\right). \tag{6}$$

Substituting this into the expression for the density matrix in the eigenbasis of $H$, we get,

$$\rho = \frac{1}{2\cosh(\beta\sqrt{1+h^2})} \begin{bmatrix} e^{-\beta\sqrt{1+h^2}} & 0 \\ 0 & e^{\beta\sqrt{1+h^2}} \end{bmatrix}. \tag{7}$$

The eigenvalues of the density matrix $\rho$ are $\lambda_1 = \frac{e^{-\beta\sqrt{1+h^2}}}{2\cosh(\beta\sqrt{1+h^2})}$ and $\lambda_2 = \frac{e^{\beta\sqrt{1+h^2}}}{2\cosh(\beta\sqrt{1+h^2})}$. Thus, the entropy is given as

$$S_2 = -\frac{e^{-\beta\sqrt{1+h^2}}}{2\cosh(\beta\sqrt{1+h^2})} \ln \frac{e^{-\beta\sqrt{1+h^2}}}{2\cosh(\beta\sqrt{1+h^2})} - \frac{e^{\beta\sqrt{1+h^2}}}{2\cosh(\beta\sqrt{1+h^2})} \ln \frac{e^{\beta\sqrt{1+h^2}}}{2\cosh(\beta\sqrt{1+h^2})}. \tag{8}$$

This expression represents the entropy of the quantum Ising system in thermal equilibrium at temperature $T$. On the other hand, a general qubit density matrix can be expanded in terms of a $2 \times 2$ identity matrix $I$, the Pauli matrices $\vec{\sigma} = (\sigma_x, \sigma_y, \sigma_z)$, and a Bloch vector $\vec{r} = (r_x, r_y, r_z)$ with $|\vec{r}| \leq 1$,

$$\rho = \frac{1}{2}(I_2 + \vec{r} \cdot \vec{\sigma}). \tag{9}$$

The eigenvalues of $\rho$ can be expressed as $\lambda_{\max} = \frac{1}{2}(1+P)$, and $\lambda_{\min} = \frac{1}{2}(1-P)$. Here $P = \lambda_{\max} - \lambda_{\min}$ is an order parameter, which in classical optical polarization is known as or the degree of polarization for a planar electromagnetic field [15,22] and the bias in quantum thermodynamics [12]. $P$ varies from 0 (completely mixed state) to 1 (pure state). Thus, the entropy expression can then be written as [5]

$$S_2(P) = -\left(\frac{1+P}{2}\right)\ln\left(\frac{1+P}{2}\right) - \left(\frac{1-P}{2}\right)\ln\left(\frac{1-P}{2}\right). \tag{10}$$

Finally, the SET $\tau_2$ from comparing Eq. (8) and Eq. (10) can be derived from $P$, which is related to the difference between the largest and the smallest population,

$$P = \lambda_{\max} - \lambda_{\min} = \tanh\left(\beta\sqrt{1+h^2}\right), \tag{11}$$

Using this relation, we solve for $\tau_2 \left(=\frac{1}{\beta}\right)$ as

$$\frac{1}{\tau_2}(\sqrt{1+h^2}) = \tanh^{-1} P = \frac{1}{2}\ln\left(\frac{1+P}{1-P}\right). \tag{12}$$

Therefore, the effective temperature is

$$\tau_2 = \frac{2\sqrt{1+h^2}}{\ln\left(\frac{1+P}{1-P}\right)}. \tag{13}$$

The system exhibits different behaviors which depend on the value of $h$, where $h = 0$ corresponds to a classical system with no quantum fluctuations, and $h = 1$ represents a strong transverse field where quantum effects play a dominant role. In the low-temperature limit (large $\beta$), the entropy approaches zero as the system tends to a pure state, whereas in the high-temperature limit (small $\beta$), the entropy reaches its maximum value.

## 2.2 SET for general finite-dimensional systems

The generalization of the SET $\tau_d$ to a $d$-dimensional density matrix may be obtained by extending the concept of spectral imbalance used in the qubit case. In a two-level system, the order parameter $P$ was defined as the difference between the largest and smallest eigenvalues of the density matrix. For a higher-dimensional system, a possible extension of this order parameter may be given by

$$P_p = \max\left(0, \lambda_1 - \sum_{i=2}^{d} \lambda_i\right), \tag{14}$$

where $\lambda_1 \geq \lambda_2 \geq \cdots \geq \lambda_d$ are the eigenvalues of the density matrix. In the limiting cases, when $P_p = 1$, the density matrix represents a pure state where one eigenvalue is 1 and the rest are 0, whereas for $P_p = 0$, the density matrix corresponds to a maximally mixed state when $\lambda_1 = \lambda_i = 1/d$. Using this formulation, the entropy-like function for bi-partitioned eigenvalues $\lambda_1$ and $\sum_{i=2}^{d} \lambda_i$ can be constructed analogously to the qubit case. The effective bi-partitioned entropy can be written as

$$S(P_p) = -\left(\frac{1+P_p}{2}\right)\ln\left(\frac{1+P_p}{2}\right) - \left(\frac{1-P_p}{2}\right)\ln\left(\frac{1-P_p}{2}\right). \tag{15}$$

This entropy-like function defined here is properly bounded between 0 and $\ln 2$, which ensures thermodynamic consistency: it reaches its maximum when the density matrix is maximally mixed and vanishes for pure states. This formulation extends the concept of effective temperature from simple two-level systems to higher-dimensions, which may provide a general framework for analyzing thermodynamic-like properties in multi-level classical and quantum systems.

For $d > 2$, there is no unique or universal way to define $h$, as a $d$-dimensional Hamiltonian can have multiple free parameters, which makes $h$ model-dependent rather than an intrinsic property of the system. To retain universality and ensure applicability across both classical and quantum systems, and independent of any specific Hamiltonian constraints, we define a generalized SET as

$$\tau_d = \frac{2}{\ln\left(\frac{1+P_p}{1-P_p}\right)}. \tag{16}$$

This formulation ensures a smooth interpolation between pure and mixed states, which makes it applicable across arbitrary dimensions.

While SET maintains universal physical bounds with the von Neumann entropy, its behavior at extreme limits presents an interesting distinction. As $\tau_d \to 0$, the von Neumann entropy naturally vanishes, which corresponds to a pure state. However, as $\tau_d \to \infty$, the system may not necessarily attain a unique maximal entropy due to variations in eigenvalue distributions across different dimensional spaces. To refine the characterization of $\tau_d$ and ensure a more general description of spectral imbalance, we propose replacing $P_p$ (Eq. (16)) with a global purity measure $P_d$ inspired by classical polarization theory [23-25]

$$P_d = \sqrt{\frac{d\gamma - 1}{d - 1}} = \sqrt{\frac{\sum_{\substack{i,j \\ i<j}}^d (\lambda_i - \lambda_j)^2}{d-1}}, \tag{17}$$

which runs from 0 at the maximally mixed distribution to 1 at a pure state, here $\gamma = \mathrm{Tr}(\rho^2)$. In terms of $d-1$ indices of purity (IPs), which provide additional structure beyond the eigenvalues and $\gamma$ [24]. $P_d$ in terms of IPs is given as [23]

$$P_d = \sqrt{\frac{d}{d-1}\left(\sum_{k=1}^{d-1} \frac{P_{(k)}^2}{k(k+1)}\right)}. \tag{18}$$

These indices are defined as [22,24]

$$P_{(k)} = \sum_{i=1}^k \lambda_i - k\lambda_{k+1}, \quad 1 \le k \le d-1, \tag{19}$$

where ordering $\lambda_1 \ge \lambda_2 \ge \cdots \ge \lambda_d$ guarantees $0 \le P_{(1)} \le P_{(2)} \le \cdots \le P_{(d-1)} \le 1$. These IPs may provide a more refined tool to characterize the purity structure of the quantum states. They can quantify the overall degree of purity or mixedness and provide the specific structure of the eigenvalue distribution of the density matrix. They offer a detailed description of how classical and quantum states are spread out across the available states in a system and represent the distribution and correlation of eigenvalues within the classical polarization space and Hilbert space. This approach is particularly valuable when we investigate the structure of quantum states in higher-dimensional systems, where the $\gamma$ may not fully incorporate the complexity of the eigenvalue distribution. Thus, the SET can be written as

$$\tau_d = \frac{2}{\ln\left(\frac{1+P_d}{1-P_d}\right)}. \tag{20}$$

Specifically, when $d = 2$, we have only one purity index $P_{(1)} = \lambda_1 - \lambda_2$ the formula for $\tau_d$ reduces to the same result as in previous works on classical polarization systems [5,15], hence, it preserves the validity of earlier formulations.

The definition of SET implies that $\tau_d \to 0$ if and only if $\rho$ is rank-1 and non-degenerate, while $\tau_d \to \infty$ is maximally mixed. Such asymptotic behavior parallels the third law in that perfectly pure states become unreachable at finite cooling steps (or finite resources), and fully disordered states sit at an infinite SET. The degenerate states, however, tend to reach at finite SET when $T \to 0$. To see this in detail we studied the behavior of $\tau_d$ at low $T$ for a specific open isotropic Heisenberg spin chain in next section.

Furthermore, von Neumann entropy in terms of IPs can be expressed as

$$S_d(\rho) = -\sum_{j=1}^{d}\left(\frac{1}{d} - \frac{P_{(j-1)}}{j} + \sum_{i=j}^{d-1}\frac{P_{(i)}}{i(i+1)}\right)\ln\left(\frac{1}{d} - \frac{P_{(j-1)}}{j} + \sum_{i=j}^{d-1}\frac{P_{(i)}}{i(i+1)}\right), \tag{21}$$

with $P_{(0)} = 0$. A convenient way to visualize these properties is through $S_d - \tau_d$ diagrams, where each density matrix $\rho$ is mapped to the point $(\tau_d(\rho), S_d(\rho))$. The diagram for $d = 2$ is a single curve which runs from $(0,0)$ at pure state to $(\infty, \ln 2)$ the maximally mixed state, whereas for higher dimensions, the space of eigenvalues is larger, and the resulting surface can exhibit rank-deficient boundary curves or cusp points. The non-analytic cusp points may echo phase-transition-like features from macroscopic thermodynamics, where, they suggest transitions in the eigenvalue distribution, such as degeneracies or vanished or zero eigenvalues.

Because the construction hinges solely on spectral data, it applies uniformly to both quantum density matrices and classical polarization coherency matrices (normalized to have unit trace). For a two-dimensional coherency matrix, $\tau_2$ reproduces the so-called effective polarization temperature [5]. Taken together, these results confirm that the finite-dimensional formulation of $\tau_d$ offers a robust analog of temperature that satisfies standard thermodynamic expectations. It requires neither uniform energy gaps or Hamiltonian specifications nor large-volume arguments; still, it can provide the thermodynamic properties such as a geometric version of the third law and aligns with familiar two-level definitions.

It is worth emphasizing that the parameter $P$, defined as the difference between the most significant and smallest eigenvalues of the density matrix, serves as a natural order parameter in the context of quantum thermodynamics and polarization theory. In classical optics, $P$ corresponds to the degree of polarization [26], while in quantum systems, it reflects the bias or spectral imbalance between energy levels [12]. This analogy positions $P$ as an indicator of distance from maximal disorder, with P = 0 representing a maximally mixed state and $P = 1$ a pure state. Consequently, the SET formula $\tau_d = 2/\ln((1 + P_d)/(1 - P_d))$ inherits a direct dependence on this order parameter, establishing a bridge between spectral asymmetry and a temperature-like quantity. This interpretation strengthens the analogy to traditional thermodynamic systems, where order parameters govern phase behavior, entropy production, and irreversibility.

## 3. Physical Interpretation of the SET

The $\tau_d$ offers a temperature-like measure for quantum states, defined via their purity $\gamma = \text{Tr}[\rho^2]$ as $P_d = \sqrt{d\gamma - 1/d - 1}$ and $\tau_d = [\tanh^{-1} P_d]^{-1}$, where $d$ is the Hilbert space dimension. Unlike the canonical temperature $T = 1/\beta$, which presupposes a Gibbs state $\rho(\beta) = e^{-\beta H}/Z(\beta)$ tied to a Hamiltonian $H$ and thermal bath, $\tau_d$ requires neither that rely only on the mixedness of states. This makes $\tau_d$ operationally meaningful in non-equilibrium or Hamiltonian-agnostic scenarios, e.g., quantum simulators or systems post-quench, where it may reflect an observable effective temperature, which controls correlation or response functions.

A practical operational window for SET emerges whenever the eigenvalue spectrum of a state remains fixed during measurement yet is directly accessible through purity. For qubit registers this can be achieved with single-copy randomized measurements: global random Cliffords, projective read-out, and correlator statistics yield $\gamma$ for subsystems of up to ten trapped-ion qubits and, in principle, 20–30 superconducting qubits [27]. When two identical copies can be prepared, SWAP-interferometry measures the same $\gamma$, already demonstrated for bosons in optical lattices with quantum-gas microscopes [28] and for ten-ion chains [29], and generalizes to qutrits and photonic two-mode couplers. In classical light fields, the density matrix is replaced by the coherency or Mueller matrix, from which simplified vectorial metrics, derived from specific element subsets, can reveal spectral asymmetry, polarization disorder, and internal optical structure; such metrics have been used to characterize spin Hall effects, layered birefringence, and even tissue pathology, and may directly yield the eigenvalue contrasts needed for computing $P_d$ and SET [30]. In all above settings—stationary spectrum, measurable purity, calibrated read-out—the SET may serve as a Hamiltonian-agnostic spectral thermometer that organizes correlation and response data just as a canonical temperature does in equilibrium thermodynamics.

The interplay between $\tau_d$ and $T$ reveals their alignment and divergence. For a Gibbs state, purity $\gamma(\beta) = Z(2\beta)/Z(\beta)^2$ maps SET to $T$. In the high-temperature limit ($\beta \to 0$), both $T$ and SET diverge proportionally, coinciding asymptotically, while at low temperatures the agreement breaks down if additional statistical entropy persists, for example due to ground-state degeneracy. Specifically, a degenerate ground-state manifold of degeneracy $g$ induces a finite plateau at zero canonical temperature,

$$\tau_d \xrightarrow[T \to 0]{} \left[\tanh^{-1}\sqrt{\frac{d/g-1}{d-1}}\right]^{-1}, \tag{22}$$

which reflects residual entropy in the degenerate subspace. We analytically and numerically studied this behavior isotropic Heisenberg chains ($\sum_{i=1}^{L-1} \vec{\sigma}_i \cdot \vec{\sigma}_{i+1}$) with open boundaries (lengths $L = 2, 3, \ldots, 9$) with a Hilbert space dimensions $d = 2^L$ and explored relation between $\tau_d$ and $T$. Numerical simulations reveal this striking parity effect in the $\tau_d$ at low temperatures ($T < 1$), as shown in Fig. 1(a) (inset). This behavior stems from the spectral properties of the Hamiltonian. For even $L$, the ground state is a unique singlet ($S = 0$), which produces a thermal state $\rho_d(\beta)$ with high purity at low $T$. As the $P_d$ increases, it drives $\tau_d$ toward zero. Conversely, odd-$L$ chains feature a degenerate $S = 1/2$ ground state due to an unpaired spin, which reduces purity and elevates $\tau_d$. At $T \to 0$, the Gibbs state for odd-$L$ is a maximally mixed projector onto a 2D subspace, while for even-$L$, it is pure. The degenerate ground states saturates at finite plateaus (e.g., $\tau_8 \approx 1.28$ for $L = 3$ diminishing slowly with increasing $L$), accurately predicted by the theoretical expression above.

At finite low $T$, this degeneracy-induced entropy persists, which causes systematically higher $\tau_d$ in odd-$L$ chains. The parity effect fades at higher $T$ as thermal fluctuations dominate. This sensitivity of $\tau_d$ to ground-state degeneracies and quantum correlations distinguishes it from conventional thermodynamic measures, which may offer insights into finite-size quantum systems. At higher temperatures ($T \gtrsim 3$), $\tau_{2^L}$ scales linearly as $\tau_{2^L} \approx A_L T$, driven by purity corrections away from the maximally mixed state. The slope $A_L = \sqrt{\frac{(2^L-1)}{3(L-1)}}$, derived from purity expansion $\gamma \approx \frac{1}{d}(1 + \text{Var}(H)\beta^2)$ (with variance $\text{Var}(H) = 3(L-1)$, see Appendix A), matches closely numerical computations (Fig. 1(b)). The exponential increase in Hilbert space dimension $2^L$ significantly accelerates purity decay, which reinforces linear scaling visually, despite underlying exponential dependencies. This behavior may represent the quantum-classical crossover at large $T$. Therefore, the $\tau_{2^L}$ robustly quantifies the mixedness of finite quantum states that reveals subtle thermodynamic effects obscured by $T$. Its universal low- and high-temperature scaling offers a sensitive probe of both degeneracy and spectral properties, which may make it a versatile diagnostic tool for probing spectral and correlation properties in quantum many-body systems, with implications for experimental and theoretical studies of low-dimensional quantum matter, quantum thermodynamics, and device characterization.

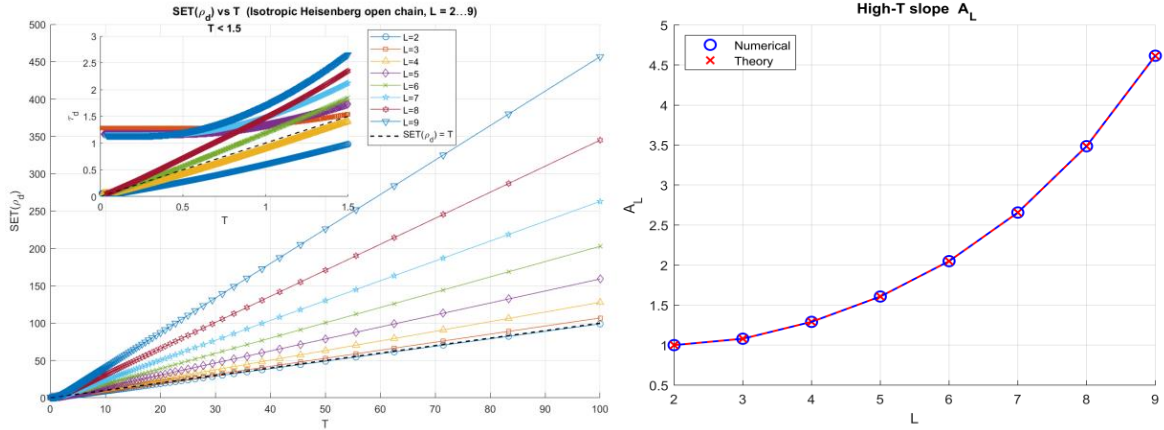

***FIG. 1 (a-b):*** *Left (a) $\tau_{2^L}$ vs canonical temperature $T$ for finite isotropic Heisenberg chains (lengths $L = 2$ to $9$). The inset highlights low-temperature plateaus for odd chains (degenerate ground states). Dashed line marks identity $T = \tau_{2^L}$. Right (b) High-temperature slope $A_L$ comparing numerical results (circles) to theoretical predictions (crosses), which verify the scaling with Hilbert space dimension.*

## 4. Entropy-SET Bounds for classical polarization states and quantum thermal states

To generate the entropy-SET diagrams, we implement a Monte Carlo algorithm to uniformly sample points on a *d*-dimensional unit sphere. Each sampled point represents a distinct set of indices of purity (IPs), $P_{(i)}, i = 1, 2, \ldots d - 1$, which determines the eigenvalue difference of the corresponding density matrix [24]. This method enables a comprehensive exploration of the parameter space, which ensures that a physically realizable space is covered. The Monte Carlo approach yields a representative and unbiased representation of the entropy-SET relationships across various dimensions by effectively sampling the space of possible density matrices. This process facilitates a detailed and accurate characterization of thermodynamic properties of a system, which may show critical features and trends in both classical and quantum regimes.

The $S_d - \tau_d$ diagram constructed using purity-based measures offers a deeper understanding of quantum state structure. While thermodynamic entropy and temperature are defined in terms of the energy distribution and specific Hamiltonian, the purity-based diagram incorporates purity and mixedness, which reveals additional structural information about quantum states. The analysis of entropy-SET relationships for different quantum systems, qubits (2D), qutrits (3D), and quartits (4D), demonstrates how the entropy behavior changes with system dimensionality. A key distinction between 2D and higher-dimensional systems is the emergence of bounded $S_d - \tau_d$ regions, as shown for the 3D and 4D cases, as opposed to a single entropy curve for the qubit case. These bounded regions are governed by the purity constraints imposed by the spectral decomposition of quantum states. The points, curves, and subregions are generated by specific constraints on the values of the IPs inequality,

$$0 \leq P_{(1)} \leq P_{(2)} \leq \cdots \leq P_{(d-2)} \leq P_{(d-1)} \leq 1, \tag{23}$$

The feasible regions of the $S_d - \tau_d$ plane for a $d \times d$ density matrix feature $d - 2$ cusp points, each corresponding to a critical transition in the entropy-SET relationship. These cusp points correspond to states where distinct IPs reach their limiting values and provide a geometric perspective on the rank structure of the Hilbert space.

The first cusp point, $A$ (cyan square in Figs. 3 and 4), is generated by setting $P_{(1)} = 0$, and $P_{(j)} = 1$ for $j = 2,3,\ldots,d-1$. The next cusp point, $B$ (red square in Fig. 4), arises from $P_{(1)} = P_{(2)} = 0$ and $P_{(j)} = 1$ for $j = 2,3,\ldots,d-1$. In general, higher-order cusp points appear as we sequentially set more of the lower-indexed IPs $P_{(k)}$ to zero, while it maintains the nested order constraint Eq. (23).

The total number of distinct $S_d - \tau_d$ curves generated by such configurations is given by the combinatorial sum,

$$(d-1) + (d-2) + \cdots + 1 = d(d-1)/2 \tag{24}$$

corresponding to all possible valid permutations of partial saturation and nullification of the IPs. For example, when $d = 4$, the inequality $0 \leq P_{(1)} \leq P_{(2)} = P_{(3)} = 1$ gives the first lower curve which is a qubit-like rank 2 curve, where $\tau_4 \to 0$ to $\tau_4$ corresponding to point $A$ (Fig 4). The second lower curve (point $A$ to point $B$) is defined by $0 = P_{(1)} \leq P_{(2)} \leq P_{(3)} = 1$, while $0 = P_{(1)} = P_{(2)} \leq P_{(3)} \leq 1$ generates the third lower curve in Fig. 4. The upper degenerate entropy curve is obtained by setting $0 \leq P_{(1)} = P_{(2)} = P_{(3)} \leq 1$. The intermediate curves $\tau_4 \to 0$ to cusp $B$ and from cusp $A$ to $\tau_4 \to \infty$ are drawn by setting $0 \leq P_{(1)} = P_{(2)} \leq P_{(3)} = 1$ and $0 = P_{(1)} \leq P_{(2)} = P_{(3)} \leq 1$, respectively.

Certain subregions in these diagrams correspond to rank-deficient regions, where the system has fewer nonzero eigenvalues. For instance, in the $d = 4$ case, the region formed by the curve connecting $\tau_4 \to 0$ to cusp point $B$, along with the two lower entropy curves, corresponds to a rank-3 region (Fig. 4) and is occupied by points following $0 \leq P_{(1)} \leq P_{(2)} \leq P_{(3)} = 1$. This structure visually provides the way entropy and SET behave in constrained purity distributions, which demonstrates how higher-rank regions are enclosed within lower-rank subspaces in both classical polarization, quantum resource theory, and quantum thermodynamics [31]. Consequently, the entropy-SET diagrams serve as powerful tools for the characterization, quantification, and classification of the states and dynamics of classical and quantum systems of arbitrary dimensions.

**4.1 The $S_2 - \tau_2$ diagram**

Since for the two-level system, the SET $\tau_2 = 1/\beta_2$ is a function of only $P_{(1)} = \tanh \beta_2$, therefore, the entropy function is given by

$$S_2 = -\left[\left(\frac{1+\tanh \beta_2}{2}\right)\ln\left(\frac{1+\tanh \beta_2}{2}\right)\right) + \left(\left(\frac{1-\tanh \beta_2}{2}\right)\ln\left(\frac{1-\tanh \beta_2}{2}\right)\right)\right], \tag{25}$$

which defines a unique curve in the $S_2 - \tau_2$ plane. This curve characterizes the evolution of entropy from 0 to $\ln d$ as the $\tau_2$ increases from zero to infinity. From the perspective of polarization theory, the black curve in Fig. 2 (marked with circles) illustrates all physically admissible entropies of a planar electromagnetic field as a function of the SET.

To provide a Hamiltonian-independent reference, we construct a family of density matrices using the Parametric Spectrum Ansatz (PSA). The PSA formalism encompasses an idealized spectral-to-entropy relationship that may apply uniformly. It provides a universal reference against which $S_2$ curves on the physically realizable space. This makes it particularly valuable in engineered quantum systems where energy-level configurations are variable but purity constraints remain intrinsic. Thus, in this approach, the eigenvalues $\mu_i$ are defined as

$$\mu_i(\zeta) = \frac{e^{-\zeta \alpha_i}}{Z(\zeta)}, \quad Z(\zeta) = \sum_{i=1}^{2} e^{-\zeta \alpha_i}, \tag{26}$$

where $\alpha_i \in \mathbb{R}$ are fixed spectral parameters, and $\zeta$ is a tunable parameter analogous to inverse temperature. This ansatz mimics a Boltzmann-like spectrum while it remains agnostic to any specific Hamiltonian dynamics. For equi-spaced energy levels $\alpha_i = [0,1]$, the eigenvalue distribution becomes geometric, while the associated purity index is given by $P_{(1)}(\zeta) = |\mu_1(\zeta) - \mu_2(\zeta)|$, with $\mu_1(\zeta) \geq \mu_2(\zeta)$, therefore, $P_2(\zeta) = P_{(1)}(\zeta)$. The resulting PSA curve $S_{PSA}(\zeta)$ appears as a smooth dashed line in Fig. 2, which lies precisely on the purity-based $S_2 - \tau_2$ diagram. This confirms the universality of the purity-based approach.

For comparison with genuine thermal states [32], consider a qubit with Hamiltonian $H = \frac{\hbar \omega}{2}\sigma_z$. The partition function is $Z = 2\cosh\left(\frac{\hbar \omega}{2T}\right)$, which leads to the thermodynamics entropy

$$S_{th}(T) = k_B\left(\ln Z - \frac{\hbar \omega}{2T}\tanh\frac{\hbar \omega}{2T}\right). \tag{27}$$

We hereafter assume $k_B = \hbar = 1$ for simplicity. For a two-level system the population imbalance under thermal statistics is

$$P_{th}(T) = \tanh\left(\frac{\omega}{2T}\right). \tag{28}$$

Whereas the purity-based framework yields

$$P_2 = \tanh\left(\frac{1}{\tau_2}\right). \tag{29}$$

Equating Eq. (28) and Eq. (29) gives a direct mapping

$$\tau_2 = \frac{2T}{\omega}. \tag{30}$$

When $\omega = 2.0$ this reduces to $\tau_d = T$, so every value of the thermodynamic temperature coincides with an identical SET; consequently $S_{th}(T)$ and $S_2$ overlap exactly (orange curve). The energy gap is tuned so that the Boltzmann population ratio produces the very same purity profile enforced by the SET construction, which makes the two entropy notions equivalent. This agreement suggests that while purity-based entropy generally extends the thermodynamic interpretation of temperature, specific choices of $\omega$ may lead to complete correspondence between the two approaches. This occurs when the entropy formulation based on purity constraints aligns with the traditional Boltzmann-Gibbs entropy formulation.

Furthermore, in Fig. 2 for small $\omega = 0.5$ and $1.0$; the entropy $S_{th}(T)$ increases rapidly with thermodynamic temperature $T$, which reaches its maximum value sooner compared to cases with larger $\omega$. This behavior arises because, for small energy level spacings, the thermal occupation probabilities tend to equalize at lower temperatures, which leads to a faster entropy saturation. Conversely, increasing the gap to $\omega = 2.5$ gives $\tau_d = 0.8T$; for a fixed $\tau_2$ one needs a higher bath temperature $T$ to achieve the same imbalance. Thermal excitation is suppressed, entropy grows more slowly, and the purple $S_{th}(T)$ curve lies below the universal purity curve. In practical terms, larger $\omega$ increases resilience of ground-state purity to thermal noise: coherence is retained to higher $T$ than the SET value alone would suggest.

The comparison therefore shows a unique resonance $\omega = 2.0$ where purity-based and Boltzmann entropies coincide, and describes how a short (enlarge) gap causes a systematic deviation, entropy saturation is at faster pace (delayed) and the purity-based SET becomes a lower (upper) envelope for the true thermodynamic entropy.

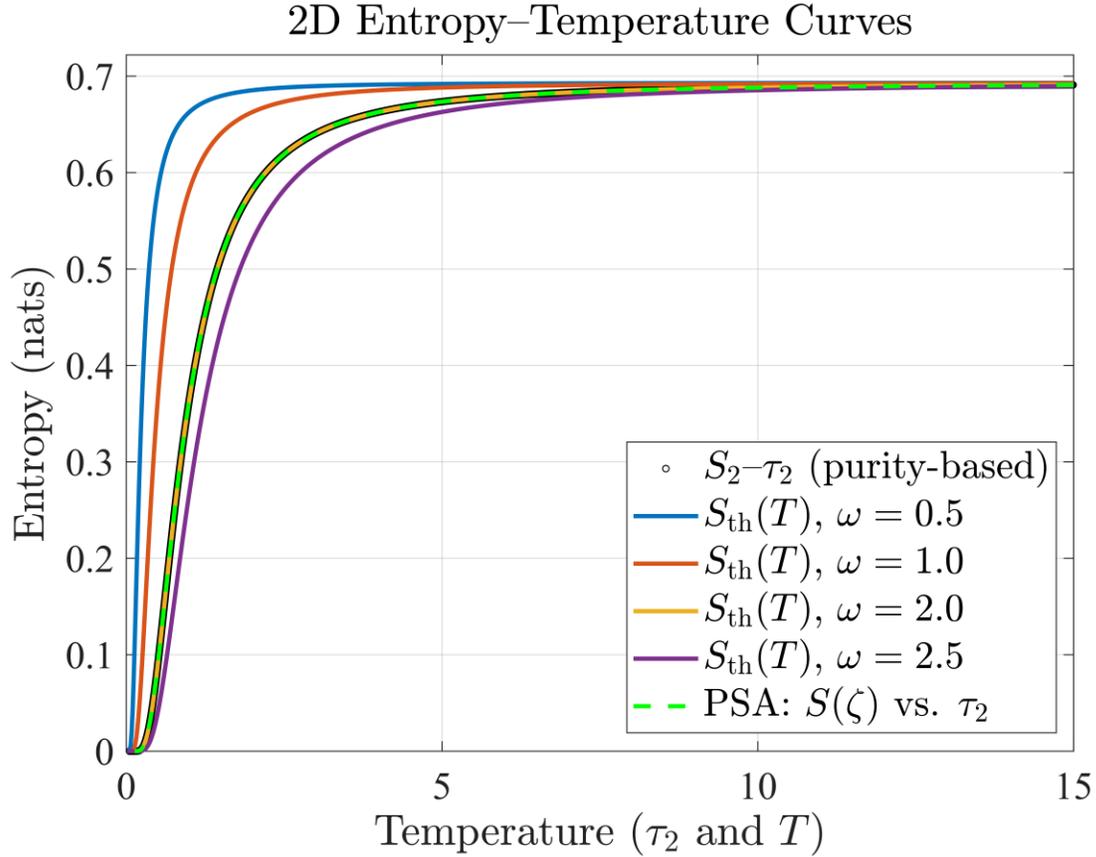

*FIG. 2 Comparison of normalized entropy as a function of temperature for different values of ω. The black curve with marker o represents entropy $S_2 - \tau_2$, derived from purity indices, while green dashed lines coinciding with $S_2$ is the $S_{PSA}(\zeta)$ curve. The colored curves represent the thermodynamic entropy $S_{th}$ as a function of standard temperature T for different Hamiltonians with ω = 0.5, 1.0, 2.0, 2.5. The plot demonstrates that, apart from scaling, the thermodynamic entropy curves follow a similar trend to the $S_2 - \tau_2$ relation, and for specific values of ω (e.g., ω = 2.0), the orange curves coincide with the $S_2$ curve.*

Our objective in this analysis is to demonstrate that the purity-based entropy-SET framework exhibits a behavior similar to the traditional thermodynamic entropy-$T$ relation, apart from differences in scaling when $\omega \neq 2$. The purity-based approach introduces an alternative way to define temperature that remains consistent with $T$ and it may offer additional insights into the purity structure of quantum states. The comparison between different values of $\omega$ reveals that while the functional form of the entropy-$T$ relation remains similar, the scaling of the $T$ varies with $\omega$. Despite these differences, the general shape of the entropy curves remains the same, which suggests that the purity-based approach provides the same fundamental thermodynamic behavior with the added advantage of information about state purity.

**4.2 The $S_3 - \tau_3$ diagram**

In contrast to the two-dimensional case, where entropy and SET are functionally related along a single curve, the $S_3 - \tau_3$ diagram for three-level systems forms a physically realizable region bounded by IPs constraints. With this approach, the SET $\tau_3$ is introduced as a spectral parameter governed by the generalized purity expression

$$P_3 = \sqrt{\frac{\left(3P_{(1)}^2 + P_{(2)}^2\right)}{4}}, \tag{31}$$

where, different purity constraints on the indices $P_{(1)}$, and $P_{(2)}$ determine the shape of the bounded $S_3 - \tau_3$ region. The entropy

$$\begin{aligned}S_3 = -\Bigg[&\left(\frac{2 + 3P_{(1)} + P_{(2)}}{6}\right) \ln\left(\frac{2 + 3P_{(1)} + P_{(2)}}{6}\right) \\ &+ \left(\frac{2 - 3P_{(1)} + P_{(2)}}{6}\right) \ln\left(\frac{2 - 3P_{(1)} + P_{(2)}}{6}\right) + \left(\frac{1 - P_{(2)}}{3}\right) \ln\left(\frac{1 - P_{(2)}}{3}\right)\Bigg].\end{aligned} \tag{32}$$

$S_3$ is now a multivalued function of $\tau_3$ and the $P_{(1)}$ [22,24]. Note that $P_{(2)} = \sqrt{4\tanh^2(1/\tau_3) - 3P_{(1)}^2}$. Thus, physically realizable $S_3 - \tau_3$ diagram is formed as shown in Fig. 3 (grey shaded region). The cyan square labeled point corresponds to $P_{(1)} = 0$ and $P_{(2)} = 1$, which implies that $\lambda_1 = \lambda_2$ and $\lambda_3 = (2\lambda_1 - 1)/2$. This spectral configuration marks a critical boundary where two eigenvalues are degenerate and dominate over the third, which suggests a phase-transition like structure in the purity space that separates different entropic regimes.

To provide a Hamiltonian-independent reference curve within this region, we consider a one-parameter family of density matrices constructed via the PSA. The eigenvalues of a qutrit density matrix are defined by

$$\mu_i(\zeta) = \frac{e^{-\zeta \alpha_i}}{Z(\zeta)}, \quad Z(\zeta) = \sum_{i=1}^{3} e^{-\zeta \alpha_i}, \tag{33}$$

Where $\alpha_i \in \mathbb{R}$ are fixed spectral parameters, and $\zeta$ is a tunable parameter analogous to inverse temperature. For equispaced spectra $\alpha_i = i - 1$, the eigenvalues form a geometric progression, which allows a closed-form expressions for entropy and purity. The PSA curve $S_{PSA}(\zeta)$, which lies entirely within the realizable $S_3 - \tau_3$ region, acts as a benchmark against which Hamiltonian-based entropic behaviors may be evaluated.

To illustrate how energy-level structure influences thermodynamic behavior, we now consider a qutrit Hamiltonian $H = \omega_1|0\rangle\langle 0| + \omega_2|1\rangle\langle 1| + \omega_3|2\rangle\langle 2|$ with $\omega_1 = 0$, $\omega_2 = 2$, and $\omega_3 = 3$, which leads to a partition function $Z = 1 + \sum_{i=2}^{3} e^{-\omega_i/T}$. The thermodynamic entropy follows the Boltzmann-Gibbs formulation

$$S_{th}(T) = -\sum_{i=1}^{3} \frac{e^{-\frac{\omega_i}{T}}}{Z} \ln \frac{e^{-\frac{\omega_i}{T}}}{Z}. \tag{34}$$

Fig. 3 provides insights into the influence of the ground-state energy level $\omega_1$ on the thermodynamic behavior of a qutrit system. By keeping the excited energy levels fixed ($\omega_2 = 2$, $\omega_3 = 3$) and varying $\omega_1$, we observe how thermal entropy evolves with $T$ for different energy-level configurations. Interestingly, the entropy curve for $\omega_1 = 0$ aligns closely with the upper boundary of the $S_3 - \tau_3$ region, which suggests that for certain energy-level configurations, thermodynamic entropy can nearly saturate the purity-based entropy constraints.

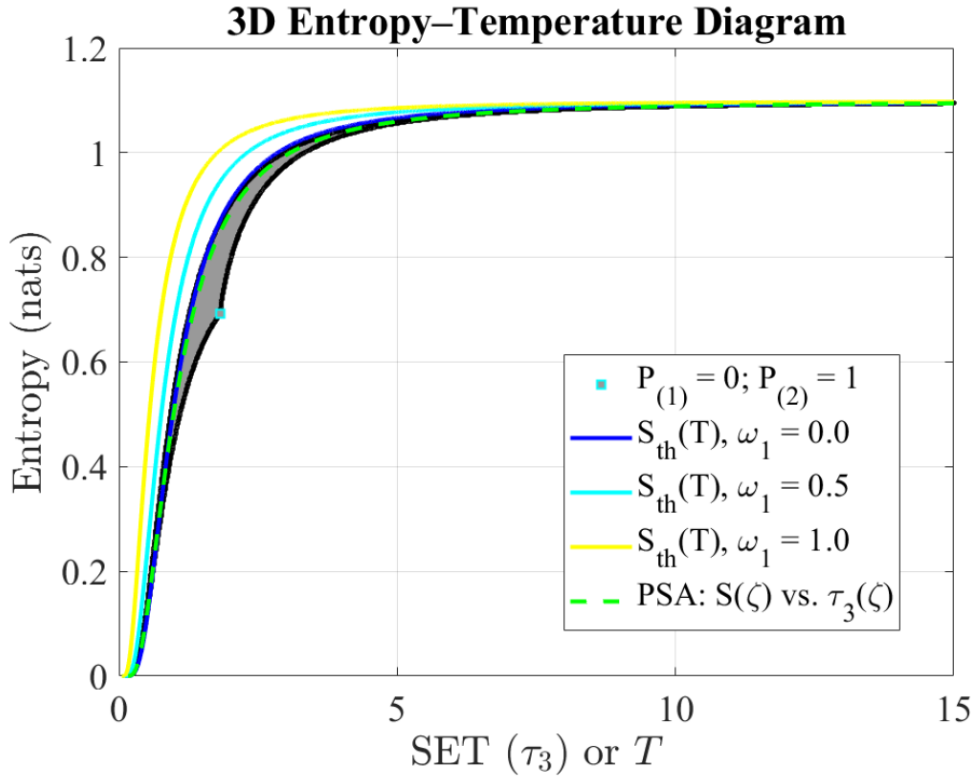

**FIG. 3** The gray-shaded region with boundary curves represents points belonging to the $S_3 - \tau_3$ diagram, derived from the two indices of purity (IPs). The green dashed curve is the PSA trajectory $S_{PSA}(\zeta)$. Solid lines show the thermodynamic entropy $S_{th}$ for $\omega_1 = 0.0, 0.5, 1.0$. Point A (cyan square) corresponds to the spectrum $\lambda_1 = \lambda_2 \gg \lambda_3$, i.e., $P_{(1)} = 0$, $P_{(2)} = 1$; it marks a degenerate-doublet corner (cusp) of the admissible region.

The increase in entropy at low temperatures with higher values of $\omega_1$ from 0.0 (blue) to 1.0 (yellow) may be interpreted as the consequence of the decrease in energy gaps between the ground and excited states, which enhances the population of higher levels even at low temperatures. This leads to a more uniform probability distribution over energy levels, which increases thermodynamic entropy. As $T$ rises, thermal fluctuations dominate, which reduces the relative impact of $\omega_1$ that causes all entropy curves to converge. Essentially, $\omega_1$ modifies the initial state occupation probabilities, which makes entropy grow faster at low temperatures while it shows a negligible effect at higher temperatures. At higher temperatures, all entropy curves $S_{th}$ (and $S_3$) tend to merge and approach the maximum entropy limit of $\ln 3$, showing thermal equilibration where all three energy levels become equally populated.

### 4.2.1 Geometric interpretations of 3D polarization states of classical electromagnetic field

We now take advantage to use $S_3 - \tau_3$ diagram to characterize and classify non-paraxial electromagnetic field. The geometric interpretations of 2D and 3D polarization states of electromagnetic field is an important issue in optics literature [33-44].

The cusp in Fig. 3 and the three curves are obtained by setting all possible constraints on the IPs inequality, $0 \leq P_{(1)} \leq P_{(2)} \leq 1$. The cusp point $A$ represents a plane unpolarized state whose polarization ellipse is completely random but remains fixed in a plane, i.e., $P_{(1)} = 0$ and $P_{(2)} = 1$ ($P_3 = 1/2$). The curve between point $A$ (excluded) and points where $\tau_3 \to 0$ represents all the possible states with $P_3 > 1/2$, i.e., $P_{(2)} = 1$ and $0 < P_{(1)} \leq 1$. These states can be considered as incoherent compositions of the two polarization eigenstates of $\rho$ with nonzero eigenvalues whose polarization planes are in general different [22]. In the limiting case that the polarization planes of both components coincide, the polarization state is called regular, otherwise, it is said to be nonregular [34].

The curve extending from point $A$ to the regions where $\tau_3 \to \infty$ corresponds to states with $P_3 < \frac{1}{2}$. These states have the first IP of zero, $P_{(1)} = 0$, which means that the two more significant eigenstates of the polarization matrix have equal weights. The second IP ranges from $0 \leq P_{(2)} \leq 1$ and with $P_{(2)} = 1$ at point $A$ represents two-component states. As $P_{(2)}$ decreases (with $P_{(2)} < 1$), the state must contain three incoherent components, eventually reaching $P_{(2)} = 0$, thus, the state can be interpreted as an equiprobable incoherent mixture of the three eigenstates of the polarization matrix. Therefore, excluding point $A$, this curve contains both regular and nonregular three-component polarization states.

At this point it is worth recalling that the degree of nonregularity of a polarization state is determined by the properties of the characteristic decomposition of the polarization density matrix [22,25], given as

$$\rho_3 = P_{(1)}\rho_{3p} + \left(P_{(2)} - P_{(1)}\right)\rho_{3m} + \left(1 - P_{(2)}\right)\rho_{3u}. \tag{35}$$

The pure component is expressed as $\rho_{3p} = (U\text{diag}(1,0,0)U^{\dagger})$, where $U$ is the unitary diagonalization matrix. It contains only the single more significant (larger eigenvalue) polarization eigenstate of $\rho_3$. The middle component $\rho_{3m} = (1/2)(U\text{diag}(1,1,0)U^{\dagger})$ is called the discriminating component, which is prepared as an equiprobable mixture of the two eigenstates with major associated eigenvalues. The arbitrary wave unpolarized component is written as $\rho_{3u} = \left(\frac{1}{3}\right)(U\text{diag}(1,1,1)U^{\dagger}) = \left(\frac{1}{3}\right)I_3$ ($I_3$, being the 3x3 indentity matrix), which has an equally probable mixture of all the three eigenstates.

The matrix $\rho_{3m}$ has always rank = 2, while the rank $m = rank(Re\rho_{3m})$ of its real part is limited by $2 \leq m \leq 3$, whose minimal value $m = rank(\rho_{3m}) = 2$ constitutes a genuine property of regular states [34].

The upper curve with maximum entropy is obtained by setting $P_{(1)} = P_{(2)}$, hence, $\rho_{3m} = 0$ in the characteristic decomposition. Therefore, the curve only contains regular polarization states without a discriminating component, consequently, a point on the curve can always be decomposed into a fully polarized component $\rho_{3p}$ and an unpolarized component $\rho_{3u}$. For all other possibilities of $P_{(1)} < P_{(2)} < 1$, the points are bounded by the three curves in which all three components are present in the characteristic decomposition.

### 4.3 The $S_4 - \tau_4$ diagram

For a four-level system, the $S_4 - \tau_4$ diagram reveals a bounded and structured region, governed by the hierarchy of IPs. From the classical polarization viewpoint, all alterations induced by a linear passive medium on an incoming 4×1 Stokes vector are encoded in its $4 \times 4$ Mueller matrix. A corresponding four-dimensional density matrix can be reconstructed from the Mueller matrix, enabling the definition of a SET $\tau_4$ with

$$P_4 = \sqrt{\frac{1}{3}\left(2P_{(1)}^2 + \frac{2P_{(2)}^2}{3} + \frac{P_{(3)}^2}{3}\right)}. \tag{36}$$

This formulation is model-independent and purely spectral, making it universally applicable to both classical and quantum systems without invoking a specific Hamiltonian. Detail universal physical relations between entropy, degree of purity, and IPs are given in Refs. [45-52].

Now introduce a PSA that provides a uni-parametric, Hamiltonian-independent framework for generating eigenvalue spectra through a Boltzmann-like form

$$\mu_i(\zeta) = \frac{e^{-\zeta\alpha_i}}{Z(\zeta)}, \alpha_i = [0,1,2,3], \tag{37}$$

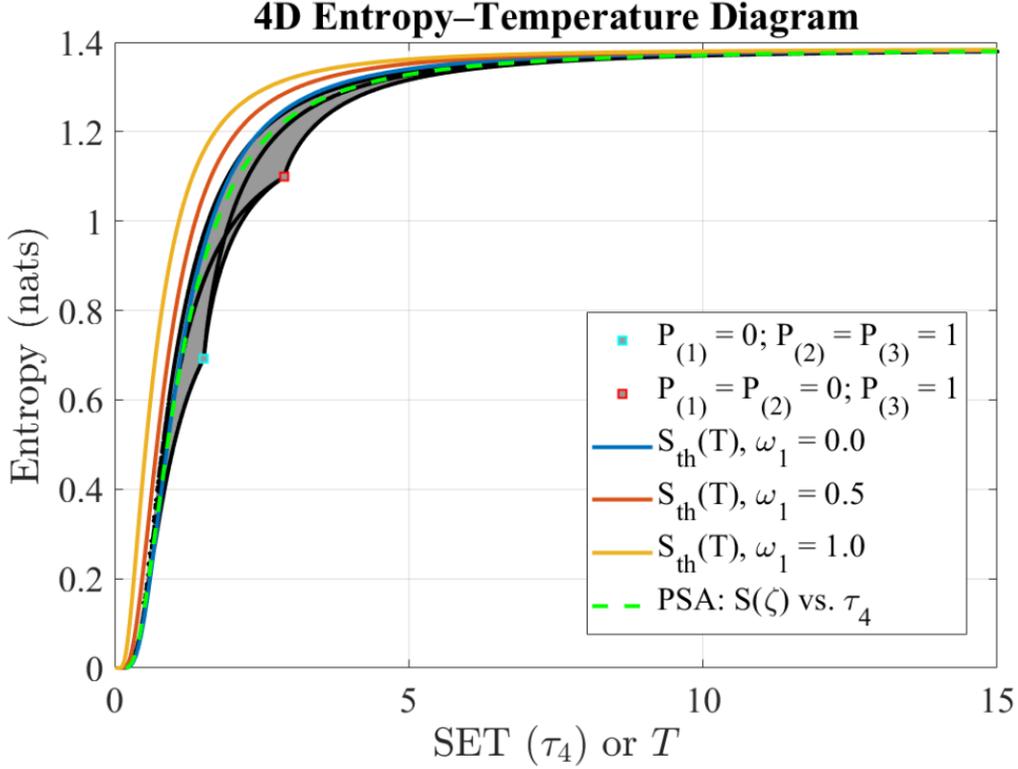

**FIG. 4** *Four-level entropy–SET diagram. The gray envelope is the set allowed by the three IPs, while solid curves give $S_{th}$ for $\omega_1 = 0.0, 0.5, 1.0$. The dashed green curve is the PSA reference $S_{PSA}(\zeta)$ vs. $\tau_4(\zeta)$. Point A (cyan square) again satisfies $\lambda_1 \geq \lambda_2 \geq \lambda_3 = \lambda_4 = 0$, i.e., $P_{(1)} = 0, P_{(2)} = P_{(3)} = 1$. Point B (red square) lies on the opposite edge with $P_{(1)} = P_{(2)} = 0, P_{(3)} = 1$ and hence $\lambda_1 = \lambda_2 = \lambda_3 > \lambda_4 = 0$. These extremal spectra illustrate how the SET bounds tighten as degeneracy patterns change.*

where $\zeta$ is a non-negative parameter that interpolates between the maximally mixed state ($\zeta \to 0$) and the pure state limit ($\zeta \to \infty$). From these eigenvalues, one computes the PSA entropy $S_4(\zeta)$ and the associated temperature $\tau_4(\zeta)$. The resulting PSA curve $S_4(\zeta)$ versus $\tau_4(\zeta)$ traces a universal, continuous path within the physically realizable entropy–SET region, independent of any Hamiltonian details. Its universality is understood as follows: (i) for every $\zeta$ the probabilities $\{\mu_i\}$ satisfy $\sum_i \mu_i = 1$ and $\mu_i \geq 0$, hence each point $(S_4(\zeta), \tau_4(\zeta))$ lies inside the entropy–SET region allowed by the four-level purity constraints; (ii) the mapping $\zeta \mapsto (S_4(\zeta), \tau_4(\zeta))$ is smooth and strictly monotone because increasing $\zeta$ simply widens or narrows the Boltzmann weight ratio without reference to any explicit level spacings. Consequently, the PSA produces a single, continuous path that is the same for all four-level systems, serving as a Hamiltonian-agnostic benchmark within the physically realizable domain.

However, from thermodynamic point of view, a quartit Hamiltonian, $H = \omega_1|0\rangle\langle 0| + \omega_2|1\rangle\langle 1| + \omega_3|2\rangle\langle 2| + \omega_4|3\rangle\langle 3|$ with $\omega_1 = 0, \omega_2 = 2, \omega_3 = 3$, and $\omega_4 = 4$ leads to a partition function $Z =$

$\sum_{i=1}^{4} e^{-\omega_i/T}$. The thermodynamic entropy $S_{th}(T) = -\sum_i p_i \ln p_i$, where the probabilities are given by $p_i = \frac{e^{-\beta\hbar\omega_i}}{Z}$. Again, the thermodynamic $S_{th}$ curve for $\omega_1 = 0$ reaches close to upper bound in the $S_4 - \tau_4$ diagram, which demonstrate that $S_{th}$ behaves as if the system were maximally mixed within the constraints set by purity. For different choices of $\omega_1$, the $S_{th}$ curves shift in a predictable manner, which depend on the specific Hamiltonian spectrum. Nonetheless, despite these variations, the overall behavior of the $S_{th}$ curves remain qualitatively similar.

The central insight the above discussion is that the purity-based entropy and SET not only encapsulate traditional thermodynamic properties such as the behaviors of $S_d$ curves in the $S_d - \tau_d$ diagrams and the third law but also provide additional information about the purity structure of quantum states.

Now it is natural to ask a question that how much useful work can be extracted from a state constrained by the same entropy or SET? To address this, we now turn to ergotropy, a fundamental resource-theoretic quantity that quantifies the extractable work via unitary operations. The aim is to establish ergotropy–entropy and ergotropy–SET bounds and demonstrate that entropy and SET serves not only as a passive state classifier but also as a powerful predictor of extractable work potential across systems with varying dimensions.

## 5. Ergotropy-Entropy and Ergotropy-SET bounds

Consider a quantum system described by a Hamiltonian given by

$$H = \sum_{i=1}^{d} \epsilon_i |\epsilon_i\rangle \langle \epsilon_i|, \tag{38}$$

where $0 \leq \epsilon_1 < \epsilon_2 < \cdots < \epsilon_d$, the ergotropy $\mathcal{W}(\rho)$ quantifies the maximum extractable work via unitary operations [53]. It is defined as

$$\mathcal{W}(\rho) = \text{Tr}(\rho H) - \text{Tr}(\rho^{\downarrow} H), \tag{39}$$

where

$$\rho^{\downarrow} = \sum_{i=1}^{d} \lambda_i |\epsilon_{d-i+1}\rangle \langle \epsilon_{d-i+1}|, \tag{40}$$

is the passive state [54] formed by aligning the eigenvalues $\lambda_1 \geq \lambda_2 \geq \cdots \geq \lambda_d$ of $\rho$ inversely with the energy levels $\epsilon_i$. This ensures $\rho^{\downarrow}$ has minimal energy for the given spectrum. The concept offers thermodynamic interpretation of quantum states, which relates it to free energy and temperature [55-57]. It provides insights for the properties of quantum batteries [58-59]. The von Neumann entropy $S(\rho)$ measures the mixedness of $\rho$, while the relative entropy of coherence $C_{\text{rel.ent}}(\rho) = S(\rho_{\text{diag}}) - S(\rho)$, where $\rho_{\text{diag}}$ is $\rho$ dephased in the energy basis, quantifies quantum coherence in $\rho$ [60]. The $C_{\text{rel.ent}}(\rho)$ quantifies the degree of quantum coherence in the energy eigenbasis by measuring the deviation of a quantum state $\rho$ from its dephased version $\rho_{\text{diag}}$, which is obtained by removing all off-diagonal elements in that basis. However, the total ergotropy can be decomposed into coherent and incoherent contributions [61], with the coherent part arises solely from superpositions in the energy eigenbasis and reflects genuinely non-classical features of the quantum state.

We consider structured quantum states characterized by ordered eigenvalues of the form

$$\lambda_1 \geq \lambda_e. \tag{41}$$

Here $\lambda_e = \lambda_2 = \lambda_3 = \cdots = \lambda_d$. These structured states provide analytically tractable bounds for ergotropy across the entire entropy domain $[0, \ln d]$ and across SET domain $[0, \infty]$. Specifically, we focus on the simplest cases, $d = 2$ and $d = 4$, as illustrative examples. The structured states $\mathcal{W}(\rho_{\text{str}})$ define a bound, which ensure that for an arbitrary quantum state $\rho$, the maximum ergotropy $\mathcal{W}_{\max}(\rho)$ for a given entropy or SET satisfies

$$\mathcal{W}_{\max}(\rho) \gtrsim \mathcal{W}(\rho_{\text{str}}) \geq \mathcal{W}(\rho) \tag{42}$$

For two-level systems with eigenvalues $\{\lambda_1, 1 - \lambda_1\}$, the left inequality in Eq. (42) becomes equality, i.e., $\mathcal{W}_{\max}(\rho) = \mathcal{W}(\rho_{\text{str}})$. This maximum ergotropy is given as

$$\mathcal{W}(\rho_{\text{str}}) = \epsilon_2 (2\lambda_1 - 1), \tag{43}$$

where $\epsilon_2$ is the excited-state energy of $H = \text{diag}(0, \epsilon_2)$. The SET bound follows from substituting $P_{(1)} = P_2 = 2\lambda_1 - 1$ into $\tau_2$.

For a four-level system with structured eigenvalues $\left\{\lambda_1, \frac{1-\lambda_1}{3}, \frac{1-\lambda_1}{3}, \frac{1-\lambda_1}{3}\right\}$ or equivalently $P_{(1)} = P_{(2)} = P_{(3)} = P_{(e)}$, the structured ergotropy explicitly becomes

$$\mathcal{W}(\rho_{\text{str}}) = \frac{\epsilon_4}{3}(4\lambda_1 - 1), \tag{44}$$

by assuming a Hamiltonian $H = \text{diag}(0, \epsilon_2, \epsilon_3, \epsilon_4)$, with $0 < \epsilon_2 < \epsilon_3 < \epsilon_4$, which is notably independent of intermediate energy eigenvalues $\epsilon_2$ and $\epsilon_3$. Moreover, the entropy and SET expressions $\rho_{\text{str}}$ states in terms of IPs are,

$$S_4(\rho) = -\frac{1}{4}\left[(1 + 3P_{(e)}) \ln\left(\frac{1 + 3P_{(e)}}{4}\right) + 3(1 - P_{(e)}) \ln\left(\frac{1 - P_{(e)}}{4}\right)\right], \tag{45}$$

and

$$\tau_4(\rho) = \frac{2}{\ln\left(\frac{\sqrt{3} + P_{(e)}}{\sqrt{3} - P_{(e)}}\right)}. \tag{46}$$

Figs. 5(a-d) illustrate numerical data clearly, which display states generated from the Ginibre and uniform-entropy sampling methods in the ergotropy–entropy and ergotropy–SET diagrams. Our numerical simulations demonstrate that while for $d = 2$, structured states define the exact ergotropy bound, for $d > 2$, these structured states yield ergotropy values $\mathcal{W}(\rho_{\text{str}})$ slightly below the absolute numerical maximum $\mathcal{W}_{\max}(\rho)$. A small fraction (less than 4%) of sampled points generated via the Ginibre ensemble lie slightly above the structured-state bound in Figs. 5(c)–(d). These outliers remain diagonal and exhibit negligible coherence, as evident from the color coding. Their deviation arises from subtle variations in eigenvalue ordering. In contrast, points from uniform-entropy sampling consistently lie below the red curve, reinforcing the structured-state bound as a tight and reliable reference for maximal ergotropy.

Coherence is quantified using the $C_{\text{rel.ent}}(\rho)$ [60]. For each randomly generated density matrix, we rotate the diagonal eigenvalue spectrum by a random unitary $Q$ which introduces coherence in the energy eigenbasis. To evaluate coherence, we first express the state in the energy eigenbasis by applying $Q^\dagger$, then remove the off-diagonal terms to obtain the dephased state. The coherence is then computed as the entropy of this dephased state minus the entropy of the original state. In Figs. 5(a-d) coherence is encoded as a color gradient and helps distinguish states with the same entropy or SET but different degrees of quantum superposition and reveals how states with similar entropy or SET can exhibit distinct internal structures and work extraction.

Since SET is a spectral quantity, it is not directly relevant to the quantum coherence. However, the relationship between SET (or entropy) and ergotropy indirectly reveals the role of coherence in work extraction. For a fixed $\tau_d$ or entropy (i.e., fixed eigenvalues), the maximum ergotropy is achieved when $\rho$ is diagonal in the energy eigenbasis of $H$, where $C_{\text{rel.ent}}(\rho) = 0$. In this configuration, the populations, the diagonal elements of $\rho$ in the energy basis, are optimally aligned, with the largest eigenvalue paired with the highest energy level. This maximizes active energy $\text{Tr}(\rho H)$, and since passive energy $\text{Tr}(\rho^\downarrow H)$ depends only on the eigenvalues, the ergotropy is maximized.

However, states with non-zero coherence exhibit superpositions that typically correlate with suboptimal population alignment, which reduce $\text{Tr}(\rho H)$. Therefore, these states have slightly lower ergotropy than the maximum for diagonal states. Conversely, for fixed $\tau_d$ or entropy, states with lower ergotropy exhibit negligible coherence. This can be interpreted as such states are typically close to the passive state $\rho^\downarrow$, which are diagonal in the energy basis and have $C_{\text{rel.ent}} \approx 0$. These nearly passive states minimize $\text{Tr}(\rho H)$ relative to the fixed $\text{Tr}(\rho^\downarrow H)$, which results in low ergotropy and minimal coherence, as opposed to states with intermediate ergotropy that often show higher coherence. This interpretation aligns with theoretical findings that establish bounds on the contribution of coherence to ergotropy through relative entropy measures $D$, specifically for the ergotropy–SET diagram, if we assume $\beta \equiv 1/\tau_d$, which emphasize that coherence enhances work extraction only within well-defined thermodynamic limits, i.e., $\beta \mathcal{W}_{\text{coh}}$ lies within the bound $|\beta \mathcal{W}_{\text{coh}} - C_{\text{rel.ent}}(\rho)| \leq D$, where $D$ quantifies the relative entropy distance to the thermal state $\rho_\beta$. Precisely, $D = D(\rho^\downarrow || \rho_\beta)$ gives the lower bound and $D = D(\Delta(\rho^\downarrow) || \rho_\beta)$ the upper bound, with $\rho^\downarrow$ and $\Delta(\rho^\downarrow)$ being passive states constructed from the original and dephased (classically diagonal) counterpart in the energy eigenbasis, respectively [61].

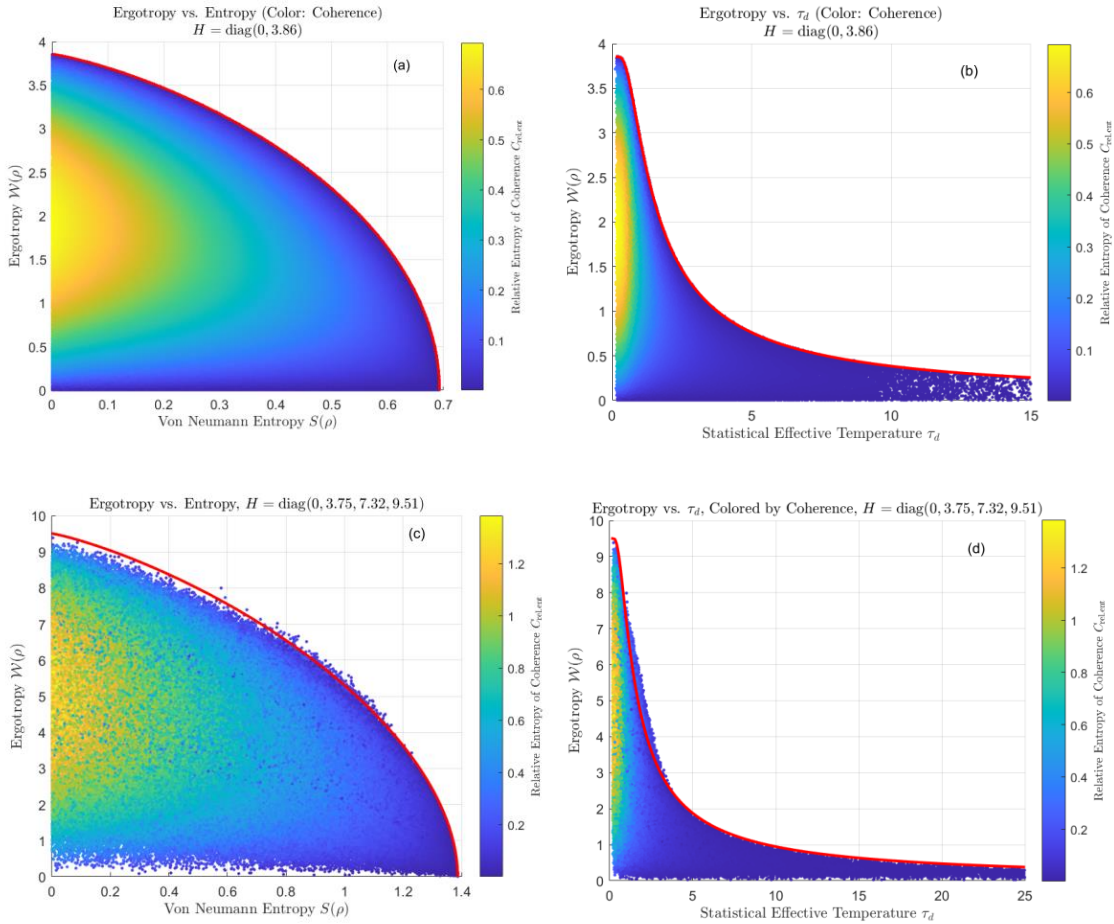

*FIG 5 (a-d). Ergotropy versus von Neumann entropy ((a) and (c)) and ergotropy versus statistical effective temperature $\tau_n$, ((b) and (d)) color-coded by the relative entropy of coherence. Top row (a) and (b) corresponds to a two-level system with H=diag(0,3.86), and bottom row (c) and (d) to a four-level system with H=diag(0,3.75,7.32,9.51) generated randomly. The red curves are derived from structured eigenvalue distributions. Coherence introduces significant variation in ergotropy for entropy and $\tau_d$ values, which may show its operational role in quantum thermodynamic performance.*

This behavior is evident in our numerical simulations, as illustrated in Figs. 5(a-d). States which achieve the maximum ergotropy for a given $\tau_d$ or entropy, those lie on (or close to) the structured states bound (red curve), exhibit near-zero coherence, which indicate they are effectively diagonal in the energy basis. It should be noted that the structured-state (red curves) bound accurately coincides with numerical maxima for qubits ($d = 2$), (see Figs. 5(a) and 5(b)) which validates the analytical equality Eq. (42). Figs. 5(c) and 5(d) show the quartits ($d = 4$) case. Here, although structure-states yield slightly lower ergotropy values compared to numerical maxima, they closely approximate the upper limit across a substantial entropy and SET range examined for a four-dimensional randomly generated diagonal Hamiltonian. The higher values of ergotropy here are only due to the contributions from incoherent source [61]. In contrast, the intermediate region is occupied by states with higher coherence (color-coded yellow) have comparatively lower ergotropy. In these states both coherent and incoherent sources may contribute to the total ergotropy [61]. At the opposite extreme, states with low ergotropy also show negligible coherence (blue points), consistent with their proximity to the passive state. Moreover, as $\tau_d \to \infty$ or the entropy $S(\rho) \to \ln d$, states approach the maximally mixed state $\rho = 1/d$. In this limit, these states are diagonal in any basis, including the energy basis, which result in $C_{\text{rel.ent}} \approx 0$. Additionally, the ergotropy $\mathcal{W}(\rho) \approx 0$ since $\text{Tr}(\rho H) \approx \text{Tr}(\rho^\downarrow H)$, as reordering them has no effect. This is reflected in Figs. 5(a-d), where at high $\tau_d$ or maximum entropy, both ergotropy and coherence vanish, with states appearing uniformly blue.

Therefore, while $\tau_d$ itself is independent of coherence, it sets the spectral potential for ergotropy, and the ergotropy-$\tau_d$ relationship demonstrates that coherence modulates the efficiency of realizing this potential. Thus, these diagrams may provide thermodynamic characterizations of quantum states. The two key benefits are: (i) they enable direct comparison across systems of different dimensionalities, and (ii) they offer meaningful assessments of quantum energetic performance without reliance on assumptions of thermal equilibrium, which is essential for practical quantum devices that operate in out-of-equilibrium regimes. Our results emphasize the utility of SET as an effective, operationally meaningful thermodynamic quantity. This approach supports precise predictions on the maximum extractable work and aligns with deeper thermodynamic principles.

In particular, the behavior of SET at extreme limits (i.e., as purity approaches unity or entropy vanishes) offers a compelling geometric and operational insight into the third law of thermodynamics. The divergence of the inverse SET in the pure non-degenerate state limit naturally connects the unattainability of absolute zero, which may provide a spectral perspective that generalizes canonical temperature concepts to arbitrary quantum systems. This motivates a closer look at how SET encodes the third law, and how purity-induced constraints govern the thermodynamic inaccessibility of ground states.

## 6. Geometric View of the Third Law of Thermodynamics

Several authors have recently attempted to interpret and quantify the third law of thermodynamics [62-65]. Notably, theorems for the limitation of the purification of resources for a full-ranked density matrix were stated [62] and the derivation for the unattainability of absolute zero temperature was determined [63]. The concept of SET may offer a unified approach to understand the unattainability of absolute zero in both classical and quantum systems. Since $\tau_d$ captures the relationship between entropy and purity, it naturally encodes the divergence of the inverse SET ($\beta_d$) as the system approaches a perfectly pure state. This divergence indicates the unattainability principle of the third law of thermodynamics. As the degree of purity $P_d$ of a $d$-dimensional density matrix approaches 1, $\beta_d$ reaches infinite, and consequently, the $\tau_d$ approaches zero. Thus, the SET may provide a spectral-based interpretation of thermodynamic limits, which may extend traditional notions of temperature and entropy to finite-dimensional systems.

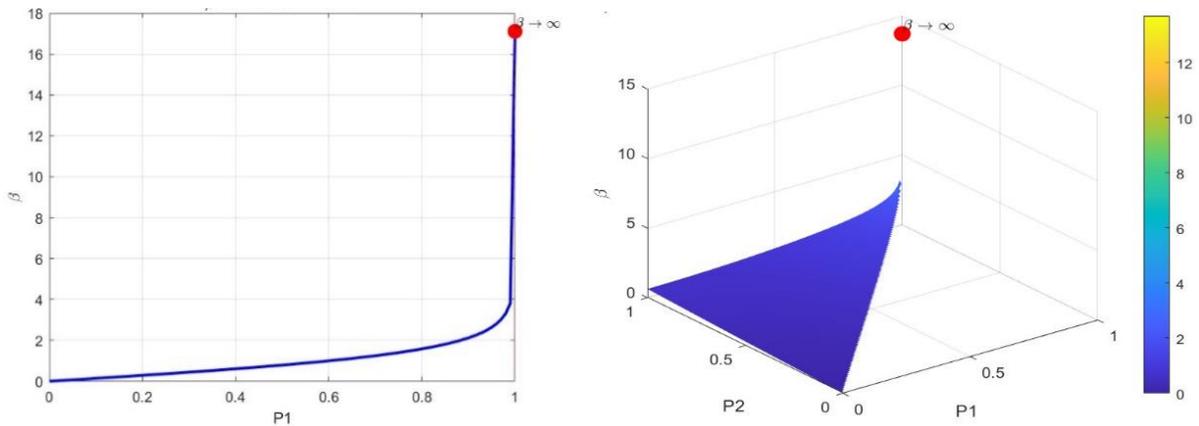

***FIG. 6(a-b)***: *(a) Inverse SET, $\beta_d$, as a function of purity for (a) $d = 2$ and (b) $d = 3$. As purity $P_d \to 1$, $\beta_d$ diverges, making the pure state an asymptotic limit. Red markers indicate the pure state.*

In Fig. 6(a), for a two-dimensional system, $\beta_2$ is plotted against $P_{(1)}$, which shows that as the system approaches a pure state $P_{(1)} \to 1$, $\beta_2$ diverges to infinity. Fig. 6(b) extends this analysis to a three-dimensional system, where $\beta_3$ is plotted against two the IPs: $P_{(1)}$ and $P_{(2)}$, constrained by $P_{(1)} \leq P_{(2)}$. The surface plot reveals a more intricate thermodynamic landscape, where different purity distributions result in distinct entropy-temperature behaviors. As both $P_{(1)}$ and $P_{(2)}$ approach 1, $\beta_3$ behaves asymptotically, which shows that complete purification becomes increasingly difficult. The red markers in both figures show the unattainable pure state, which reinforces the notion that absolute zero is an idealized limit that cannot be reached through any finite process.

Moreover, a complete pure state, whether classical or quantum, corresponds to a purity $P_{(1)} = 1$, and this purity condition is closely tied to the idealized SET of zero. Mathematically, this is obtained by the inverse SET $\beta_d$ and its relation to the IPs,

$$\beta_d = \frac{1}{2}\ln\left[\frac{1+\sqrt{\frac{d}{d-1}\left(\sum_{k=1}^{d-1}\frac{P_{(k)}^2}{k(k+1)}\right)}}{1-\sqrt{\frac{d}{d-1}\left(\sum_{k=1}^{d-1}\frac{P_{(k)}^2}{k(k+1)}\right)}}\right], \tag{47}$$

If the state is maximally pure, all the IPs are equal to 1, i.e.,

$$P_{(1)} = P_{(2)} = \cdots = P_{(d-1)} = 1, \tag{48}$$

This leads to,

$$\sum_{k=1}^{d-1}\frac{P_{(k)}^2}{k(k+1)} = \sum_{k=1}^{d-1}\frac{1}{k(k+1)}. \tag{49}$$

Using the telescoping identity

$$\sum_{k=1}^{d-1}\frac{1}{k(k+1)} = 1 - \frac{1}{d}. \tag{50}$$

Thus,

$$\sqrt{\frac{d}{d-1}\left(\sum_{k=1}^{d-1}\frac{P_{(k)}^2}{k(k+1)}\right)} = \sqrt{\frac{d}{d-1}\left(1-\frac{1}{d}\right)}. \tag{51}$$

This simplifies to

$$\sqrt{\frac{d}{d-1}\left(1-\frac{1}{d}\right)} \approx 1. \tag{52}$$

Thus, the inverse SET $\beta_d$ diverges logarithmically as the state approaches maximal purity $P_{(1)} \to 1$, which indicates the asymptotic inaccessibility of perfect purity at any finite temperature. Equivalently, if we invert this relation, the SET $\tau_d \to 0$ only in the limit. This establishes a purity-based formulation of the third law of thermodynamics that to achieve $\tau_d = 0$ requires $\beta_d = \infty$, which is physically unattainable. This aligns with the third law's unattainability principle which demonstrates that absolute purity (i.e., a pure state with zero entropy) can only be achieved in the limit of infinite inverse SET, or equivalently, at zero SET.

Furthermore, as the system cools and $\tau_d \to 0$, the qubit curves corresponding to purity distributions $0 \leq P_{(1)} \leq P_{(2)} = \cdots = P_{(d-1)} = 1$ merges with the uppermost boundary of the $S_d - \tau_d$ diagram. The upper boundary represents states of maximal entropy, where all IPs are equal, therefore Eq. (47) and Eq. (48) ensure the entropy-SET slope,

$$\frac{\partial S(P_{(k)})}{\partial \tau_d} \to 0. \tag{53}$$

The slope flattens out completely, which reinforces that absolute zero remains unattainable in any finite process. This confirms that, near absolute zero, further purification becomes increasingly difficult and thermodynamically costly. Hence, the third law of thermodynamics is geometrically represented in the SET framework, which demonstrate that the closer a system comes to a pure state, the more thermodynamically expensive it becomes to achieve further purification. This geometric representation of the third law underscores the growing thermodynamic cost of purification, whether in classical polarization theory or quantum thermodynamics, and establishes a direct link between spectral properties of the density matrix and physical resource constraints in cooling processes.

This geometric viewpoint offers a deep understanding of the connection between SET, entropy, and purity, and provides a visual and geometric interpretation of the third law. It also establishes a direct link between the spectral properties of the density matrix and the physical resources required to manipulate these systems, which shows the universal nature of the unattainability principle across classical and quantum domains.

## 7. Discussion

In this work, we introduce the statistical effective temperature (SET) as a general approach to describe thermodynamic-like behavior in finite-dimensional quantum and classical systems. We define SET in terms of IPs, which characterize the eigenvalue distribution of a density matrix. Our approach provides a universal temperature-like parameter that remains independent of specific energy gaps or equilibrium conditions. This flexibility allows SET to be applied across a broad range of systems, which includes classical polarization and quantum thermal states, without the constraints typically imposed by conventional definitions of temperature [5-13].

The comparison of SET to the canonical temperature $T$ in finite isotropic Heisenberg chains (up to $L = 9$) reveals a striking parity effect of low-temperature-plateaux in odd-length chains due to ground-state degeneracy. While the canonical temperature $T \to 0$ in these cases, the corresponding $\tau_d$ remains finite, which describes how SET encodes spectral disorder independently of energy-level scaling and shows its distinct sensitivity to degeneracy structure.

One of the significant outcomes of this study is to establish universal entropy-SET ($S_d - \tau_d$) relations in terms of indices of purity (IPs). The $d-1$ independent IPs not only quantify mixedness but also encode structural details about the eigenvalue distribution, which directly impact the thermodynamic properties of a system. In classical polarization optics, the indices of purity (IPs) serve to classify polarization states according to their purity structure and degree of disorder [24-25, 33-34, 37-39, 45-48]. In quantum thermodynamics, a similar role is played by spectral descriptors such as population bias and normalization, which are used to characterize virtual qubits and define thermal behavior within autonomous machines [12]. These quantities provide a foundation for understanding coherence, population inversion, and thermalization in finite-dimensional systems [12]. Therefore, we employ a spectral-based IPs dependent approach to unify different thermodynamic interpretations and provide a generalized $S_d - \tau_d$ framework that bridges quantum and classical domains.

To explore the space of physically realizable quantum states, we introduce a parametric spectrum ansatz (PSA). The ansatz assigns eigenvalues through a thermal-like law applied to a fixed set of dimensionless energy levels. Unlike approaches that depend on a specific qubit Hamiltonian [16], the PSA provides a deterministic and model-independent method from the maximally mixed state to a pure state, and every point generated by PSA lie within the $S_d - \tau_d$ diagram. Although variational ansätze appear widely in quantum statistical mechanics and thermal-state approximations [16-19], the present construction relies solely on spectral data constrained by purity and entropy. This simplicity suits spectrum-based thermodynamic analyses. The PSA also admits a clear thermodynamic reading: for a fixed set $\{\alpha_i\}$, the parameter $\zeta$ behaves as an effective inverse SET, so each spectrum matches an effective Gibbs-state. Consequently, every PSA trajectory occupies the $S_d - \tau_d$ region and traces thermal-like paths where purity and entropy vary coherently. This may confirm that the diagram encloses both arbitrary mixed states and ensembles that coincide with effective Gibbs distributions, which illustrates the universal reach of the SET framework.

Besides general theoretical insights, we also examine Hamiltonian-specific entropy-temperature behaviors for qubit (2D), qutrit (3D), and quartit (4D) thermal states, which reveal that, apart from a scaling factor, the thermal entropy curves obtained from canonical temperature follow trends similar to those in the $S_d - \tau_d$ diagram. Notably, for specific values of the energy gap parameter $\omega_1$, the thermodynamic entropy curves align closely with the upper boundary. This supports the universality of our approach. This connection between SET and standard thermodynamic entropy-temperature suggests that our approach provides a spectral-based interpretation of temperature, which may offer additional structural insights beyond effective temperature definitions.

While the SET faithfully captures spectral disorder through the normalized purity $P_d$, it is blind to basis-dependent coherences. Hence a diagonal mixture $\rho = \text{diag}(p_i)$ and its coherent conjugate $\hat{\rho} = U\rho U^\dagger$ (with non-diagonal $U$) share the same SET even though they differ operationally in work extraction, metrology, and phase locking [66-68]. To reveal such phase-sensitive resources one can pair SET with a coherence monotone $C(\rho)$, for instance the $l_1$-norm of coherence $C_{l_1}(\rho) = \sum_{i\neq j}|\rho_{ij}| - 1$ or the relative entropy of coherence $C_{\text{rel.ent}}(\rho)$, where the latter is a good coherence measure under incoherent operations (IO) [60,69]. The joint diagnostics (SET, $C_{\text{rel.ent}}(\rho)$) may disentangle thermodynamic resources: SET captures eigenvalue-driven entropy and any ground-state degeneracy, while $C_{\text{rel.ent}}(\rho)$ ignores that degeneracy yet reveals work-relevant coherences within it. The pair can provide a comprehensive picture of disorder and quantum phase structure, essential for far-from-equilibrium systems. Crucially, SET remains responsive to ground-state degeneracy, which alters spectral weights and hence purity, as demonstrated in our analysis of open Heisenberg spin chains, where SET ($T \to 0$) correctly captures the degeneracy-induced disorder plateau.

As a measure of spectral purity, $\tau_d$ sets the potential for work extraction, and the ergotropy–SET relation reveals how coherence influences the realization of this potential. Maximal ergotropy at fixed $\tau_d$ is achieved only when coherence vanishes, which align the state optimally with the Hamiltonian, while low ergotropy states are similarly incoherent due to their passive-like character. In the limit $\tau_d \to \infty$, both coherence and ergotropy vanish, which indicate convergence to a maximally mixed state. Thus, SET not only bounds extractable work but also exposes the operational role of coherence through the structure of optimal and passive states.

A particularly important contribution of this work is the geometric interpretation of the third law of thermodynamics within the $S_d - \tau_d$ diagram. Unlike the energy–entropy diagram used in quantum thermodynamics, which plots the von Neumann entropy $S(\rho)$ against the average energy $E(\rho) = \text{Tr}(H\rho)$ for a fixed Hamiltonian [70], the SET approach defines a basis-independent temperature-like parameter solely from the eigenvalue spectrum of the density matrix. In our formulation, the unattainability principle naturally emerges as a consequence of the divergence of inverse SET in a finite-dimensional Hilbert space constraints. As the system approaches a pure state (rank-1 density matrix), the entropy–SET slope flattens as $\tau_d \to 0$, which may indicate an increased thermodynamic cost of a state purification. This behavior directly shows the unattainability principle of the third law, which emphasizes that progressively lower SET require exponentially increasing resources. The geometric structure of the $S_d - \tau_d$ diagram may provide a visual and quantitative manifestation of this fundamental thermodynamic limitation.

Despite its broad applicability, our work also has certain limitations, primarily related to its focus on finite-dimensional systems, it cannot not fully encompass the behavior of large-scale thermodynamic systems approaching the infinite limit. In addition, SET formulation inherently depends on the spectral purity and eigenvalue distributions of finite-dimensional states. This dependence implicitly assumes reliable access to complete spectral information. In practical experimental scenarios, complete spectral characterization can be challenging due to measurement imperfections, noise, and limited tomography precision. Such practical constraints could restrict the accuracy and applicability of SET-based thermodynamic characterizations. Moreover, it does not explicitly address dynamical aspects, such as state evolution, decoherence, or thermalization processes. These dynamical effects are crucial for a complete understanding of thermodynamics, particularly in quantum regimes where open-system interactions, environmental coupling, and non-Markovian effects significantly influence the thermodynamic behavior. Although the PSA provides a systematic approach to exploring state space, it assumes a thermal-like functional form and a fixed set of dimensionless energy-like parameters. Thus, the generality of this ansatz may overlook other physically relevant spectral families or alternative eigenvalue distributions that deviate significantly from the chosen thermal-inspired form. This simplification may limit our description's capacity to fully represent more complex or highly structured quantum states appearing in certain practical systems, such as quantum many-body states exhibiting multipartite entanglement or strongly correlated structures. Furthermore, many important collective phenomena, such as critical behavior, non-classical correlations, and large-volume effects, require additional considerations that may be beyond the scope of finite-dimensional purity-based SET. To extend this work to incorporate continuous-variable systems or the thermodynamic limit remains an open area for future work.

Nevertheless, the results presented here lay the foundation for a unified theory of entropy, ergotropy, and SET that bridges classical and quantum thermodynamics. Our approach, based on the SET formulation in terms of IPs, retains significant thermodynamic features, which include low temperature degeneracy detection, the third law consistency, entropy, ergotropy bounds. This broad applicability paves the way for further research into geometric and spectral interpretations of thermodynamic principles, particularly in the study of thermal fluctuations and wave-based phenomena in both classical and quantum domains.

## 8. Conclusion

We introduce the statistical effective temperature (SET), a spectrum-based functional that does not rely on any Hamiltonian. Our formulation applies to every finite-dimensional classical or quantum system. We express SET and entropy through purity (and the indices of purity), which gives a universal $S_d - \tau_d$ landscape that unites polarization optics with quantum thermodynamics and reveals rank-dependent boundaries, cusp points, and transitions that echo phase changes. The same geometry interprets the third-law unattainability principle: as the spectrum moves toward purity, the inverse SET diverges and the entropy slope flattens for non-degenerate states, so each further purification step demands disproportionate resources. We also establish explicit connections between SET, entropy, and ergotropy, which provide operational bounds that quantify work extraction across different finite-dimensional systems, independent of thermal equilibrium assumptions. This work may set the foundation for a spectrum-based approach to thermodynamic behavior across different physical settings.

**Acknowledgment**

This work was supported by the National Natural Science Foundation of China (Grant Nos. 12475009, 12075001, and 61601002), Anhui Provincial Key Research and Development Plan (Grant No. 2022b13020004), Anhui Province Science and Technology Innovation Project (Grant No. 202423r06050004), and Anhui Provincial University Scientific Research Major Project (Grant No. 2024AH040008). T.A. would like to acknowledge Mirpur University of Science and Technology for granting ex-country leave.

**Appendix A: The variance Var($H$)**

The variance Var($H$) for the isotropic Heisenberg Hamiltonian $H = \sum_{i=1}^{L-1} h_i$, with $\vec{\sigma}_i \cdot \vec{\sigma}_{i+1}$ at temperature $T \to \infty$, where $L$ is the number of spins in an open chain is $3(L-1)$. Here, $\rho \approx \mathbb{I}/d$, $d = 2^L$, so $\langle A \rangle = \text{Tr}(A)/d$. The expectation values $\langle H \rangle = 0$, $\text{Tr}(h_i) = 0$ as Pauli matrices are traceless. Therefore,

$$\langle H^2 \rangle - \langle H \rangle^2 = \langle H^2 \rangle = \frac{1}{d} \text{Tr}\left( \sum_i h_i^2 + \sum_{i \neq j} h_i h_j \right). \tag{A1}$$

Proof: To compute $\langle h_i^2 \rangle$, we first derive the identity $h_i^2 = 3I - 2h_i$. For two spins, $h_i = \vec{\sigma}_1 \cdot \vec{\sigma}_2$, the triplet states ($S = 1$, degeneracy 3) has the eigenvalue $\lambda_{\text{triplet}} = +1$, while for singlet state ($S = 0$, degeneracy 1) has the eigenvalue $\lambda_{\text{singlet}} = -3$. Using projectors $P_{\text{triplet}}$ and $P_{\text{singlet}}$, $h_i = P_{\text{triplet}} - 3P_{\text{singlet}}$ and $P_{\text{triplet}} + P_{\text{singlet}} = 1$.

Solving, $P_{\text{singlet}} = (I - h_i)/4$, $P_{\text{triplet}} = (3I + h_i)/4$. Then, $h_i^2 = P_{\text{triplet}} + 9 P_{\text{singlet}} = [(3I + h_i) + 9(I - h_i)]/4 = 3I - 2h_i$.

However, for the full chain, since $h_i$ acts on spins $i$ and $i+1$, $\text{Tr}(h_i^2) = \text{Tr}_{i,i+1}(h_i^2) \cdot 2^{L-2} = \text{Tr}_{i,i+1}(3I - 2h_i) \cdot 2^{L-2}$. On two spins, $\text{Tr}(3I) = 12$, $\text{Tr}(2h_i) = 0$, so $\text{Tr}(h_i^2) = 12 \cdot 2^{L-2} = 3 \cdot 2^L$. Therefore, $\langle h_i^2 \rangle = 3 \cdot 2^L / 2^L = 3$.

Consider two cases for $i \neq j$: (i) non-adjacent bonds ($|i-j| \geq 2$), here $h_i$ and $h_j$ act on disjoint spins, so $\text{Tr}(h_i h_j) = \text{Tr}(h_i)\text{Tr}(h_j) = 0$; (ii) adjacent bonds ($|i-j| = 1$), for $h_i h_{i+1} = (\vec{\sigma}_i \cdot \vec{\sigma}_{i+1}) \cdot (\vec{\sigma}_{i+1} \cdot \vec{\sigma}_{i+2})$, expand into Pauli products such that,

$$(\vec{\sigma} \cdot \vec{\sigma})^2 = \sum_{l,m} \sigma_i^l \sigma_{i+1}^l \sigma_{i+1}^m \sigma_{i+2}^m = \sum_{l,m} \sigma_i^l (\delta_{lm}\mathbb{I} + i\epsilon_{lmn}\sigma_{i+1}^n)^2 \sigma_{i+2}^m. \tag{A2}$$

The trace vanishes due to the odd number of Pauli operators in each term and, $\langle h_i h_j \rangle = 0, \forall\, i = j$. Finally, $\text{Var}(H) = \sum_{i=1}^{L-1} \langle h_i^2 \rangle = \sum_{i=1}^{L-1} 3 = 3(L-1)$, which completes the proof.

## Appendix B: Structured States and Their Ergotropy

Structured states serve as analytically tractable reference states for bounding the ergotropy of quantum states at fixed spectral quantities, such as entropy or SET ($\tau_d$). They maximize population alignment with the Hamiltonian while remaining diagonal in the energy eigenbasis, and show negligible coherence contributions in ergotropy.

### B.1 Definition of Structured States

Consider a quantum state $\rho_d \in \mathcal{D}(\mathbb{C}^d)$ with spectrum

$$\lambda = \{\lambda_1, \lambda_e, \ldots, \lambda_e,\}, \quad \lambda_e = \frac{1-\lambda_1}{d-1}, \quad \lambda_1 \in \left[\frac{1}{d}, 1\right]. \tag{B1.1}$$

For a non-degenerate Hamiltonian $H = \sum_{i=1}^{d} \epsilon_i |\epsilon_i\rangle\langle\epsilon_i|$ with $0 = \epsilon_1 < \epsilon_2 < \cdots < \epsilon_d$, the structured state is

$$\rho_{\text{str}} = \lambda_1 |\epsilon_d\rangle\langle\epsilon_d| + \sum_{i=1}^{d-1} \lambda_e |\epsilon_i\rangle\langle\epsilon_i|. \tag{B1.2}$$

This maximizes $\text{Tr}[\rho H]$ by aligning $\lambda_1$ with $\epsilon_d$ and $\lambda_e$ with lower energies, while the passive state $\rho^\downarrow$ ensures minimal energy and zero coherence.

### B.2 Theorem: Ergotropy of Structured States

The ergotropy of $\rho_{\text{str}}$ is

$$\mathcal{W}(\rho_{\text{str}}) = \epsilon_d(\lambda_1 - \lambda_e) = \epsilon_d \left(\frac{d\lambda_1 - 1}{d-1}\right). \tag{B2.1}$$

Proof: Active energy is $\text{Tr}[\rho_{\text{str}} H] = \lambda_1 \epsilon_d + \lambda_e \sum_{i=1}^{d-1} \epsilon_i$. Passive energy is $\text{Tr}[\rho_{\text{str}}^\downarrow H] = \lambda_1 \epsilon_1 + \lambda_e \sum_{i=2}^{d} \epsilon_i$. Therefore,

$$\mathcal{W}(\rho_{\text{str}}) = \lambda_1 \epsilon_d + \lambda_e \sum_{i=1}^{d-1} \epsilon_i - \lambda_1 \epsilon_1 - \lambda_e \sum_{i=2}^{d} \epsilon_i = (\lambda_1 - \lambda_e)(\epsilon_d - \epsilon_1). \tag{B2.2}$$

Setting $\epsilon_1 = 0$, and $\lambda_e = \frac{1-\lambda_1}{d-1}$, we obtain $\mathcal{W}(\rho_{\text{str}}) = \epsilon_d \left(\frac{d\lambda_1 - 1}{d-1}\right)$.

### B. 3 Maximality and Numerical Observations

For $d = 2$, $\rho_{\text{str}}$ achieves the global maximum ergotropy. For $d > 2$, it forms a tight upper envelope in ergotropy-SET (ergotropy-entropy) diagrams (e.g., Figs. 5(a-d)), with states exceeding it by less than 3% in rare cases, per numerical simulations for 2 million Ginibre and uniform samples matrices.

### B.4 Zero Coherence of Structured States

Claim: $\rho_{\text{str}}$ has $C_{\text{rel.ent}}(\rho) = 0$]

Proof: Since $\rho_{\text{str}}$ is diagonal in $\{|\epsilon_i\rangle\}$, $\Delta\rho_{\text{str}} = \rho_{\text{str}}$, where $\Delta$ dephases in the energy basis. Therefore, $C_{\text{rel.ent}}(\rho) = 0$. By the Schur-Horn theorem, equality of diagonal elements to eigenvalues confirms diagonality.

Hence, structured states offer a coherence-free, near-optimal reference for ergotropy, which emphasize spectral alignment over coherence in work extraction, as validated by simulations (Figs. 5(a-d)). This may bridge spectral thermodynamics and quantum coherence for quantum battery design.